\documentclass{article}
\usepackage{ijcai19}

\usepackage{times}
\usepackage{soul}
\usepackage{url}
\usepackage[hidelinks]{hyperref}
\usepackage[utf8]{inputenc}
\usepackage[small]{caption}
\usepackage{graphicx}
\usepackage{amsmath}
\usepackage{booktabs}
\usepackage{algorithm}
\usepackage[noend]{algpseudocode}
\title{Attributed Network Embedding for Incomplete Attributed Networks}

\author{
Chengbin Hou$^{1,2}$ \and
Shan He$^2$ \and
Ke Tang$^1$
\affiliations
$^1$Southern University of Science and Technology~~~ $^2$University of Birmingham\\
\emails
chengbin.hou10@foxmail.com~~~~ s.he@cs.bham.ac.uk~~~~ tangk3@sustech.edu.cn
}

\begin{document}

\maketitle

\begin{abstract}
    Attributed networks are ubiquitous since a network often comes with auxiliary attribute information e.g. a social network with user profiles. Attributed Network Embedding (ANE) has recently attracted considerable attention, which aims to learn unified low dimensional node embeddings while preserving both structural and attribute information. The resulting node embeddings can then facilitate various network downstream tasks e.g. link prediction. Although there are several ANE methods, most of them cannot deal with incomplete attributed networks with missing links and/or missing node attributes, which often occur in real-world scenarios. To address this issue, we propose a robust ANE method, the general idea of which is to reconstruct a unified denser network by fusing two sources of information for information enhancement, and then employ a random walks based network embedding method for learning node embeddings. The experiments of link prediction, node classification, visualization, and parameter sensitivity analysis on six real-world datasets validate the effectiveness of our method to incomplete attributed networks.
\end{abstract}

\section{Introduction}
A network/graph, which consists of a set of nodes/vertices and links/edges, is a widely used data representation. In the real-world scenarios, it often comes with auxiliary/side information \cite{cai2018comprehensive,cui2018survey,hamilton2017representation}. An \textit{attributed network} can naturally include such auxiliary information as node attributes to better describe complex systems \cite{liao2018attributed,gao2018deep,huang2017accelerated}. For example, for a citation network, one may transform paper title into its attributes using NLP techniques \cite{pan2016tri}; for a social network, one may transform user profiles into its attributes using one-hot encoding \cite{liao2018attributed}; and even for a pure network, one may encode node degrees as its attributes \cite{kipf2017semi}.

Network Embedding (NE) a.k.a. Network Representation Learning has become an emerging topic in Data Mining, Machine Learning, and Network Science \cite{goyal2018graph,cai2018comprehensive,hamilton2017representation}. Typically, NE aims to learn low dimensional node embeddings while preserving one or more network properties \cite{cai2018comprehensive}. The resulting node embeddings\footnote{Essentially, node embeddings are just the data points in a low dimensional vector space, so that the off-the-shelf distance metrics and Machine Learning techniques can be easily applied.} can then facilitate various network downstream analytic tasks \cite{cui2018survey} such as link prediction \cite{lu2011link,wei2017cross,liao2018attributed} and node classification \cite{huang2017accelerated,yang2015network,hamilton2017inductive}.

There have been many successful Pure-structure based Network Embedding (PNE) methods \cite{perozzi2014deepwalk,tang2015line,cao2015grarep,wang2016structural,grover2016node2vec,ou2016asymmetric}. However, PNE methods cannot utilize widely accessible attribute information, which is highly correlated with structural information (so called homophily) \cite{tsur2012s,mcpherson2001birds}. Attributed Network Embedding (ANE), which aims to learn unified low dimensional node embeddings while preserving both structural and attribute information, has recently attracted considerable attention \cite{pan2016tri,huang2017accelerated,liao2018attributed}.

\subsection{Incomplete Attributed Networks}
Nevertheless, most of existing ANE methods have not considered the incomplete attributed networks with missing links and/or missing node attributes.

The \textit{incomplete structural information} i.e. missing links can be observed in many real-world networks: in a social network, some abnormal users e.g. criminals may intentionally hide their friendships, and some newly registered users may have none or very limited friendships; in a terrorist-attack network where each node denotes an attack and two linked attacks are committed by the same organization, it is well-known that many anonymous attacks are not clearly resolved yet \cite{lin2012community}; and so on and so forth.

The \textit{incomplete attribute information} i.e. missing node attributes may exist in some real-world networks e.g. in a social network, many users nowadays are unwilling to provide personal information due to worrying about personal privacy. Furthermore, it becomes harder to crawl complete attributed networks due to the development of anti-crawler techniques, especially while crawling data from the word-leading companies such as Facebook and Tencent.

\subsection{The Challenges}
The incomplete attributed networks bring several challenges to existing NE methods. Firstly, most PNE methods such as DeepWalk \cite{perozzi2014deepwalk} and Node2Vec \cite{grover2016node2vec}, may obtain less accurate node embeddings, because they can only utilize (incomplete) structural information. Secondly, some ANE methods such as GCN \cite{kipf2017semi} and SAGE \cite{hamilton2017inductive}, relied on links to aggregate node attributes, are likely to fail especially for those nodes with none or few links. Thirdly, the ANE methods based on matrix factorization such as TADW \cite{yang2015network} and AANE \cite{huang2017accelerated} may not converge due to factorizing over too sparse matrix caused by missing links and/or missing attributes. And finally, the ANE methods based on dense deep neural networks like ASNE \cite{liao2018attributed} may lack training samples due to missing links, since they require links to build training samples.

To sum up, the existing methods have not considered incomplete attributed networks, and hence, they do not have the mechanism to compensate missing information. 

\subsection{Our Idea}
\begin{figure*}[htbp]
    \setlength{\abovecaptionskip}{0.05cm}
    \setlength{\belowcaptionskip}{0cm} 
    \centering
    \includegraphics[width=0.95\textwidth]{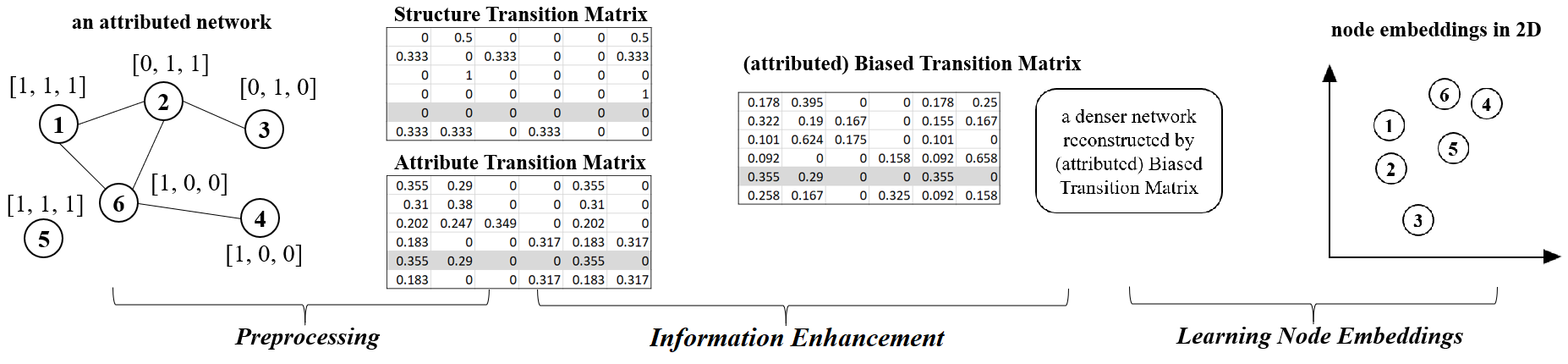}
    \caption{The illustration of ABRW method: I. obtain two transition matrices according to structural and attribute information respectively (note the demo preserves the top-3 attribute similar nodes in Attributed Transition Matrix); II. calculate Biased Transition Matrix by fusing the two matrices (note the demo assumes structure and attribute are equally important); III. learn node embeddings based on the reconstructed denser network using a weighted random walks based PNE method.}
    \label{f1}
\end{figure*}

To tackle the challenges, a mechanism is designed to compensate incomplete structural information with available (but may also be incomplete) attribute information, and vise versa. In general, our idea is to 1) reconstruct a \textit{unified denser} network in which all nodes gain much richer relationships by fusing two sources of information via transition matrices for information enhancement; 2) employ a weighted random walks based PNE method to learn node embeddings based on the reconstructed network. In particular, the information enhancement step is designed to compensate missing information with each other. The proposed method, Attributed Biased Random Walks (ABRW), is illustrated in Figure \ref{f1}.

\subsection{Contributions}
The contributions are summarized as follows: 
\begin{itemize}
    \item We justify and investigate a largely ignored real-world problem of embedding incomplete attributed networks.
    \item We propose an ANE method for incomplete attributed networks by learning embeddings on the reconstructed denser network after information enhancement. Several experiments show that our method consistently outperforms the state-of-the-art methods in most cases.
    \item An open-source framework including several network embedding methods is available at \textit{\url{https://github.com/houchengbin/OpenANE}} for benefiting future research and industrial applications.
\end{itemize}

\section{Notations and Definitions}
Let $\mathcal{G}=(\mathcal{V},\mathcal{E},W,A)$ be a given attributed network where $\mathcal{V}=\{v_1,\dots,v_n\}$ denotes a set of $|\mathcal{V}|$ nodes; $\mathcal{E}=\{e_{ij}\}$ denotes a set $|\mathcal{E}|$ links; the weight associated to each link is a scalar $w_{ij}$; the attributes associated to each node are in a row vector $A_i$;  $i,j\in \{1,\dots,n\}$ are the subscripts. Note that the proposed method can accept either directed or undirected and either weighted or unweighted attributed networks.

\paragraph{Definition 1.} Structural Information Matrix $W\in\mathcal{R}^{n\times n}$: The structural information refers to network linkage information, which is encoded in matrix $W$. There are several popular choices to encode structural information \cite{ou2016asymmetric} such as the first order proximity that gives the information of immediate/one-hop neighbors of a node, the second order proximity that gives the information of two-hop neighbors of a node, etc. In this work, the first order proximity is used to define $W$ a.k.a. the adjacency matrix.

\paragraph{Definition 2.} Attribute Information Matrix $A\in\mathcal{R}^{n\times m}$: The attribute information refers to network auxiliary information associated with each node, which is encoded in matrix $A$ where each row $A_i\in\mathcal{R}^m$ corresponds to the node attribute information for node $v_i$. To obtain the vector presentation $A_i$, one may employ word embedding technique if it is textural auxiliary information \cite{pan2016tri}, and one may employ one-hot encoding technique if it is categorical auxiliary information \cite{liao2018attributed}.

\paragraph{Definition 3.} Attributed Network Embedding: It aims to find a mapping function $Z=f(W,A)$ where $Z\in\mathcal{R}^{n\times d}$ and each row vector $Z_i\in\mathcal{R}^d$ is the node embedding vector. The pairwise similarity of node embeddings should reflect the pairwise similarity of the nodes in original attributed network.

\section{The Proposed Method}
\subsection{Preprocessing}
A \textit{transition matrix} a.k.a. Markov matrix or stochastic matrix is a square matrix where each entry is a nonnegative real number. And in this work, each row of transition matrix gives discrete probability distribution $\pi_i\in\mathcal{R}^n$ to indicate the probability of a walker to the next node from node $v_i$.

\textbf{Structural Transition Matrix} $T^W\in\mathcal{R}^{n\times n}$: This matrix is used to sample the next node from current node $v_i$ based on the discrete probability distribution given by row vector $T_i^W$ i.e. $i^{th}$ row of $T^W$. To calculate $T^W$, we have:
\begin{equation}
    T_{i,j}^W=f_{rowNorm}(W)=\frac{W_{i,j}}{\sum_{j\in n} W_{i,j}}
\end{equation}
where $f_{rowNorm}$ is a function operating on each row of $W$ such that each row becomes a probability distribution. Note that, the structural transition matrix might not be a strict transition matrix, since the isolated node leads to all-zero row. One may assign the uniform distribution to those rows, nevertheless, we retain all-zero rows to \textit{avoid the meaningless} (or misleading) links in the later reconstructed network. 

\textbf{Attribute Similarity Matrix} $S^A\in\mathcal{R}^{n\times n}$: This matrix stores the similarity measurements of attribute information between every pair of nodes in a network. Recall that the given attribute information is $A\in\mathcal{R}^{n\times m}$ where each row $A_i\in\mathcal{R}^m$ corresponds to the node attribute information for node $v_i$. To calculate $S^A$, we have:
\begin{equation}
    S_{i,j}^A=f_{similarity}(A_i, A_j) \overset{\underset{\mathrm{def}}{}}{=} \frac{A_i A_j^T}{|A_i| |A_j|}
\end{equation}
where $f_{similarity}$ is a function to measure the similarity between every pair of rows of $A$. In this work, we adopt cosine similarity as the measure. Previous work \cite{strehl2000impact} has shown that cosine similarity is a good measure in both continuous and binary vector spaces, although one may try other similarity measure e.g. Jaccard similarity.

\textbf{A sparse operator} $\Theta(\cdot)$: The aim of this operator is to make attribute similarity matrix $S^A$ sparser. In practice, $S^A$ is often a very dense matrix with few zeros, which would lead to a high computational cost in subsequent sampling stage. Our sparse operator is defined as:
\begin{equation}
    S_{i,j}^A=\Theta(S^A)=\left\{
    \begin{array}{lcl}
    0         & {if\ S_{i,j}^A < top\ k\ of\ S_i^A}\\
    S_{i,j}^A & {otherwise}
    \end{array} \right.
\end{equation}
where the sparse operator $\Theta(\cdot)$ operates on each row of $S^A$, so as to preserve the largest $k$ values but set the rest to zeros. This can \textit{avoid building links for those dissimilar} (or not such similar) nodes in the later reconstructed network.

\textbf{Attribute Transition Matrix} $T^A\in\mathcal{R}^{n\times n}$: It is similarly defined as the structural transition matrix $T^W$ as shown in Eq. (1), except it is from attribute information perspective.

\subsection{Information Enhancement}
In order to compensate the incomplete information with each other, we obtain a biased transition matrix by fusing the above two transition matrices for information enhancement, which results in a denser network with much richer relationships.

\textbf{Biased Transition Matrix} $T\in\mathcal{R}^{n\times n}$: This transition matrix aims to fuse two sources of information. The information enhancement is achieved by the following equation:
\begin{equation}
    T_i=\left\{
    \begin{array}{lcl}
    T_i^A &{if\ T_i^W\ is\ all\ zeros}\\
    \alpha T_i^W+{(1-\alpha)T}_i^A &{otherwise}
    \end{array} \right.
\end{equation}
where $T_i$, $T_i^W$ and $T_i^A$ are the $i^{th}$ rows of the corresponding transition matrices. In cases of isolated nodes, the row vector $T_i^W$ is all zeros and we directly assign attribute information $T_i^A$ to $T_i$ for compensation. For other cases, we apply a balancing factor $\alpha$ to trade-off two sources of information. 

\textbf{The reconstructed network}: The reconstructed network is then established based on the biased transition matrix $T$ after information enhancement. And it comes with several properties: 1) it reflects both structural and attribute information; 2) it is a weighted and directed network, which is more informative than an unweighted and undirected network; and 3) it does not contain isolated nodes and each node in the reconstructed network gains much richer relationships.

\subsection{Learning Node Embeddings} \label{section24}
The problem is now transformed into learning node embeddings on the reconstructed network (without attributes). There have been many successful PNE methods. But considering scalability, we follow random walks based PNE methods e.g. DeepWalk to learn node embeddings. Because of the weighted reconstructed network, differently from DeepWalk, we apply \textit{weighted random walks} to each node in a network, so as to generate a list of node sequences. 

For each node sequence, a fixed-size window is used to slide along it. For each window, several training pairs $(n_i, n_j)$ are generated such that $n_i$ is the center node and $n_j\in \mathcal{N}_i$ is the remaining/neighboring nodes. By doing so, we obtain a list of training pairs $\mathcal{D}$ for all sequences, which is then fed into Skip-Gram Negative Sampling (SGNS) model \cite{mikolov2013distributed} for training node embeddings. For each pair $(n_i, n_j)$, we \textit{maximize} the following objective function:
\begin{equation}
    log\sigma (Z_i\cdot Z_j) + m\cdot \mathbf{E}_{n_k\sim P_{\mathcal{D}}}[log\sigma (-Z_i\cdot Z_k)]
\label{eq5}
\end{equation}
where $\sigma$ is the Sigmoid function, $Z_i$ is the node embedding vector for node $n_i$, $m$ is the number of negative samples, and $n_k$ is the negative sample from the unigram distribution $P_{\mathcal{D}}$ \cite{levy2014neural}. The aim of maximizing Eq. (\ref{eq5}) is to make embedding vectors \textit{similar} if they co-occur, and \textit{dissimilar} if they are negative samples.

The overall objective is to sum over all $(n_i,n_j) \in \mathcal{D}$ i.e. $\sum_{n_i\in \mathcal{V}} \sum_{n_j\in \mathcal{N}_i} \#(n_i,n_j) Eq. (\ref{eq5})$. Intuitively, the more frequently a pair of nodes co-occurs, the more similar they are.


\subsection{Algorithm Implementation}
For better reproducibility and understanding, we summarize the core implementation details in Algorithm \ref{alg1}.

\begin{algorithm}[htbp]
\caption{Attributed Biased Random Walks}
\label{alg1}
\textbf{Input}: structural information matrix $W$, attribute information matrix $A$, top-k value $k$, balancing factor $\alpha$, walks per node $r$, walk length $l$ \\ \textbf{Output}: a list $walks$
\begin{algorithmic}[1] 
\State Compute $T^W$, $S^A$, $\Theta(S^A)$, and $T^A$ sequentially according to Eq. (1), (2), and (3)
\State Obtain biased transition matrix $T$ according to Eq. (4)
\State Initialize a list $walks=[~]$
\For{$iter\in r$ }
    \For{$v_i\in\mathcal{V}$ }
        \State Initialize a list $walk=[v_i]$
        \For{$walk\_iter\in l$ }
            \State $curr\_node=walk[-1]$
            \State $prob\_dt = T_{curr}$
            \State $next\_node = AliasSampling(prob\_dt)$
            \State Append $next\_node$ to $walk$
        \EndFor
        \State Append $walk$ to $walks$
    \EndFor
\EndFor
\State \textbf{return} a list $walks$
\end{algorithmic}
\end{algorithm}

To save memory usage and further reduce time complexity while generating a list of walks, as shown in lines 9 and 10, we directly adopt the corresponding probability distribution based on biased transition matrix $T$, so as to \textit{avoid explicitly reconstructing} the network; and then, we employ alias sampling method to efficiently simulate a random walk.

Once we obtain a list of walks/sequences by Algorithm \ref{alg1}, we follow \cite{perozzi2014deepwalk} to employ the well-developed Python library \textit{Gensim} \cite{gensim2010} and its efficient API Word2Vec for learning node embeddings. According to Section \ref{section24}, some key parameters used in the API are: model SGNS, window size $w=10$, negative samples $m=5$, and the exponent used in negative sampling distribution $3/4$.

\subsection{Complexity Analysis}

Regarding algorithm \ref{alg1}, for lines 1 and 2, the time consuming operations are in Eq. (3), which aims to seek the top-k most similar nodes for a node. Instead of fully sorting all $|\mathcal{V}|$ elements, we employ the \textit{introselect} algorithm to find the element in the top-k position without fully sorting other elements, which has average speed $O(1)$ and worse case performance $O(|\mathcal{V}|)$. Besides, for lines 3-13, the overall complexity is $O(r|\mathcal{V}|l)$ and note that, alias sampling in line 10 only requires $O(1)$ time \cite{grover2016node2vec}. Algorithm \ref{alg1} finally returns $r|\mathcal{V}|$ walks. The sliding window with length $w$ along each walk with length $l$ gives $(l-w+1)(w-1)$ training pairs, and the overall pairs are $r|\mathcal{V}|(l-w+1)(w-1)=|\mathcal{D}|$. To train node embeddings, we maximize Eq. (\ref{eq5}) by feeding all training pairs. The complexity for each pair is $O(1+m)$ and the overall complexity is $O((1+m)|\mathcal{D}|)$.

\section{Experiments}
The attributed networks tested in the experiments are summarized in table \ref{t1}. MIT, Stanford, and UIllinois are three Facebook \textit{social networks} for each university, and there are seven properties associated with each Facebook user: status flag, gender, major, second major, dorm, year, and high school \cite{traud2012social}. We take out "year" as classes, and the remaining six properties are converted into attributes using one-hot encoding. The missing values are encoded by all-zero. For the \textit{citation networks}, we use the data preprocessed by \cite{yang2016revisiting}: the attributes for Cora and Citeseer are in binary vectors, but are in continuous vectors for Pubmed.

\begin{table}[htbp]
\setlength{\abovecaptionskip}{0.15cm}
\setlength{\belowcaptionskip}{0cm}
\centering
\begin{tabular}{llllll}
\hline
Datasets   & Nodes  & Links   & Attributes & Classes  \\
\hline
MIT    	   & 6402  & 251230  & 2804       & 32   \\ 
Stanford   & 11586 & 568309  & 3306       & 37   \\ 
UIllinois  & 30795 & 1264421 & 2921       & 34   \\
Citeseer   & 3327  & 4732    & 3703       & 6    \\ 
Cora       & 2708  & 5429    & 1433       & 7    \\ 
Pubmed     & 19717 & 44338   & 500        & 3    \\ 
\hline
\end{tabular}
\caption{The summary of datasets used in the experiments}
\label{t1}
\end{table}

To simulate missing links i.e. incomplete structural information, we \textit{randomly} remove a certain percentage of links for each of the six networks. To also investigate missing attributes i.e. incomplete attribute information, we introduce the three social networks with 
\textit{inherent} missing attributes.


\subsection{Baseline Methods and Settings}
All the methods compared in the experiments are in \textit{unsupervised} fashion i.e. no label is required during embedding.

\begin{itemize}
    \item DeepWalk \cite{perozzi2014deepwalk}: It is one of the most successful PNE methods based on random walks, which considers \textit{only structural information}. 
    \item AttrPure: It considers \textit{only attribute information} by applying SVD for dimensionality reduction on the attributed similarity matrix as introduced in Eq (2). 
    \item TADW \cite{yang2015network}: It jointly models attribute and structural information as a bi-convex optimization problem under the framework of matrix factorization.
    \item AANE \cite{huang2017accelerated}: It is similar to TADW, but the problem is solved in a distributed manner.
    \item SAGE-Mean \cite{hamilton2017inductive}: The idea is to aggregate attribute information from the neighboring nodes and then, take the element-wise mean over them. For fair comparison, we adopt its unsupervised version.
    \item SAGE-GCN: GCN was first proposed by \cite{kipf2017semi}, which is designed for semi-supervised learning. For fair comparison, we adopt the unsupervised generalized GCN by \cite{hamilton2017inductive}.
\end{itemize}


We adopt the original source code of TADW, AANE, SAGE-GCN and SAGE-Mean. For hyper-parameters, we follow the suggestions by the original papers: 1) for all methods, node embedding dimension $d=128$; 2) for DeepWalk and ABRW, walks per node $r=10$, walk length $l=80$, window size $w=10$, and top-k value $k=30$; 3) for TADW, AANE and ABRW, the balancing factors are set to 0.2, 0.05 and 0.8 respectively; 4) for SAGE-GCN and SAGE-Mean, learning rate, dropout rate, batch size, normalization, weight decaying rate, and epochs are set to [search: 0.01, 0.001, 0.0001], 0.5, 128, true, 0.001, and 100 respectively. We repeat the experiments for ten times and report their averages.



\subsection{Link Prediction}
\begin{figure}[htbp]
    \setlength{\abovecaptionskip}{0.1cm}
    \setlength{\belowcaptionskip}{0cm}
    \centering
    \includegraphics[width=0.5\textwidth]{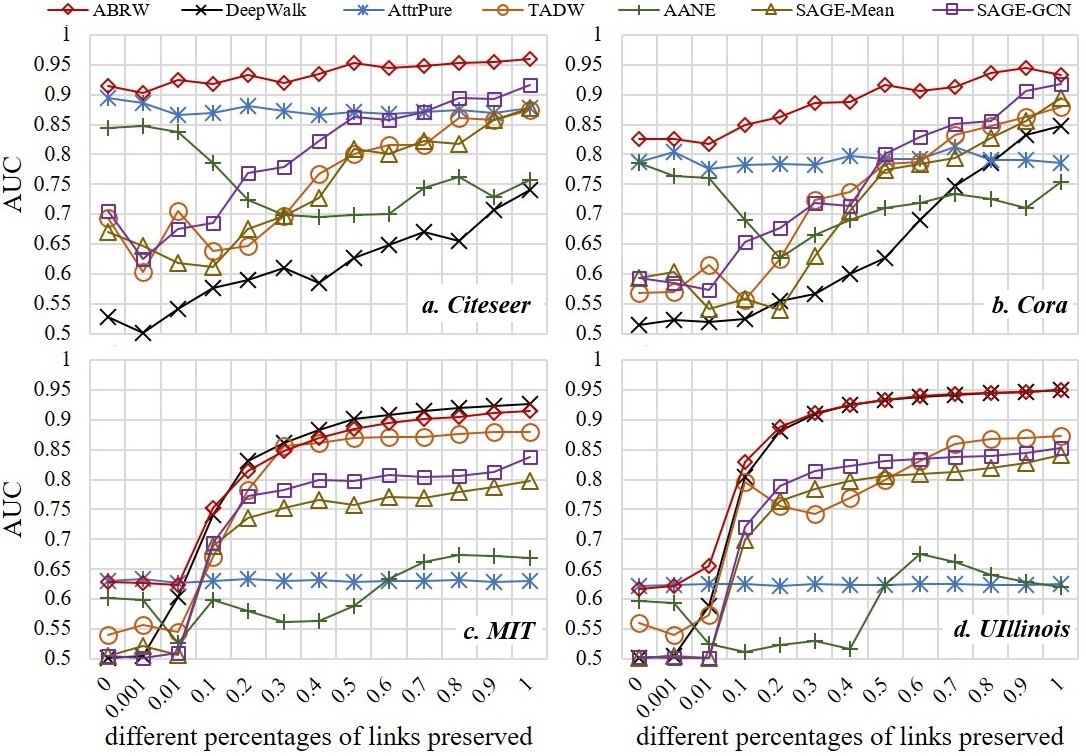}
    \caption{Link prediction task on two citation networks (Citeseer and Cora) and two social networks (MIT and UIllinois): The X-axis shows different percentages of links preserved for embedding. The Y-axis indicates AUC scores ranging [0.5, 1].}
    \label{f2}
\end{figure}

Link prediction task is intended to predict missing links or potential links. We randomly pre-remove $10\%$ links as positive samples, and generate the equal number of non-existing links as negative samples, and finally, all samples serve as \textit{ground truth}. We further randomly remove different percentages of links and use the remaining links while learning embeddings. We employ cosine similarity as the measure to predict links and report AUC scores as shown in Figure \ref{f2}.

For two citation networks Citeseer and Cora, our method outperforms all baseline methods for all different percentages of links preserved. DeepWalk receives the worst results due to that it cannot utilize attribute information, which however, is helpful for Citeseer and Cora, since we observe the impressive results by AttrPure. For all ANE methods except AANE (may not converge sometimes), they get better results when the available structural information is increasing.

For two social networks MIT and UIllinois, surprisingly, DeepWalk on MIT receives the best results when the percentages of links beyond $20\%$, but our method obtains the best results (or equally good results as DeepWalk) for all other cases on MIT and UIllinois. We can explain such contrast results of DeepWalk on the social networks w.r.t. citation networks from two aspects: 1) The social networks have far more links than that on the citation networks as shown in Table \ref{t1}; 2) The social networks have inherent missing attributes as mentioned above, which leads to less helpful attribute information and hence, the degenerated performances of ANE methods. Note that the results of AttrPure on the social networks are much worse than that on the citation networks.


\subsection{Node Classification}
\begin{figure}[htbp]
    \setlength{\abovecaptionskip}{0.1cm}
    \setlength{\belowcaptionskip}{0cm}
    \centering
    \includegraphics[width=0.5\textwidth]{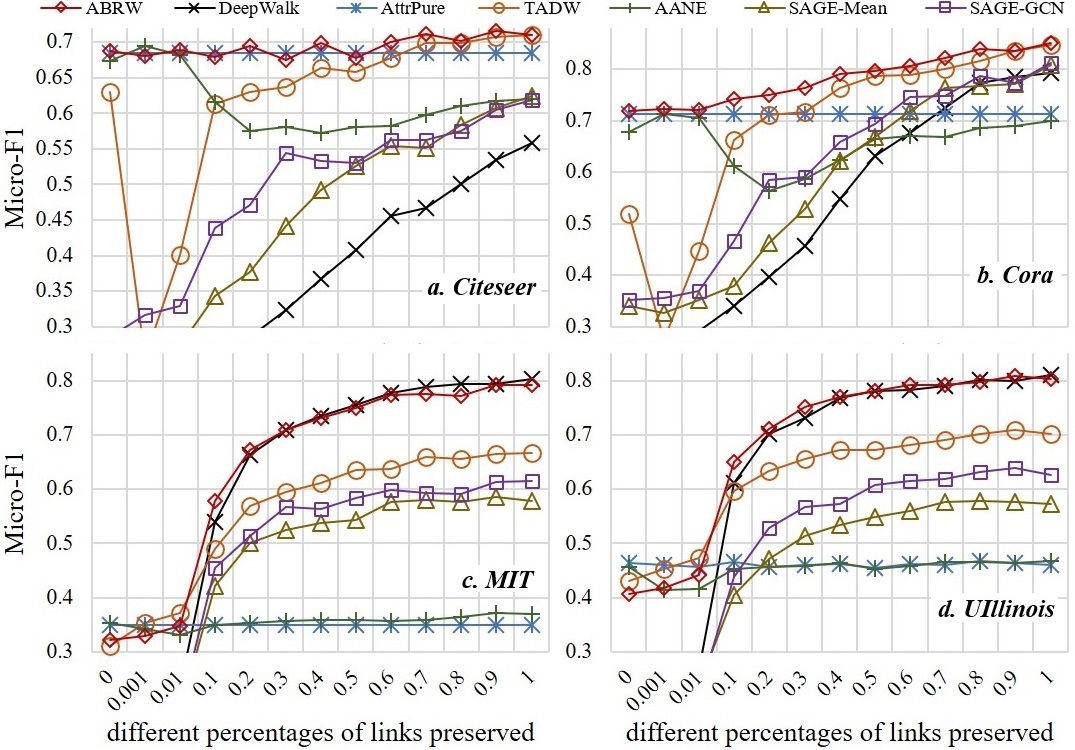}
    \caption{Node classification task on two citation networks (Citeseer and Cora) and two social networks (MIT and UIllinois): The Y-axis indicates Micro-F1 scores starting from 0.3 (hence some points are not shown) and ending near the highest score.}
    \label{f3}
\end{figure}

Node classification task is intended to assign the existing labels to the nodes without labels e.g. classify social network users into groups for precision marketing. We randomly pick $50\%$ nodes with labels for training a classifier and then, the remaining ones serve as \textit{ground truth}. Besides, we also randomly remove different percentages of links for embedding. We take one-vs-rest Logistic Regression as the classifier and report micro-F1 scores as shown in Figure \ref{f3}.

For Citeseer, attribute information dominates structural information, since AttrPure significantly outperforms DeepWalk. Even in such extreme case, our method can obtain superior results for most cases, which implies our method can robustly utilize attribute information, even if the structural information is not such helpful when only few links preserved for embedding. For Cora, our method receives the best results for all cases, and TADW, SAGE-GCN and SAGE-Mean obtain comparable results when there are sufficient links. Moreover, TADW on Citeseer and Cora drops sharply when the percentage of links is around $0.1\%$ due to factorizing too sparse structure information matrix leads to divergence.

For two social networks, the general tendency and findings are similar to link prediction task on the same datasets. Note that our method outperforms DeepWalk with a large margin for the percentages of links below $10\%$, since our method can utilize attribute information for compensating the highly incomplete structural information.


\subsection{2D Visualization}
Visualization task further reduces the dimensionality of node embeddings to 2D by PCA, and then assigns different colors to nodes according to their labels. This task paves a way to intuitively explain why the resulting node embeddings can benefit downstream tasks. Here we show the 2D visualization results trained on MIT with randomly $50\%$ missing links and inherently missing attributes. MIT attributed social network has 32 classes/years as shown in Table \ref{t1}, and there are 4932 non-isolated users (out of 6402) from year 2004 to 2009.

\begin{figure}[htbp]
    \setlength{\abovecaptionskip}{0cm}
    \setlength{\belowcaptionskip}{0cm} 
    \centering
    \includegraphics[width=0.50\textwidth]{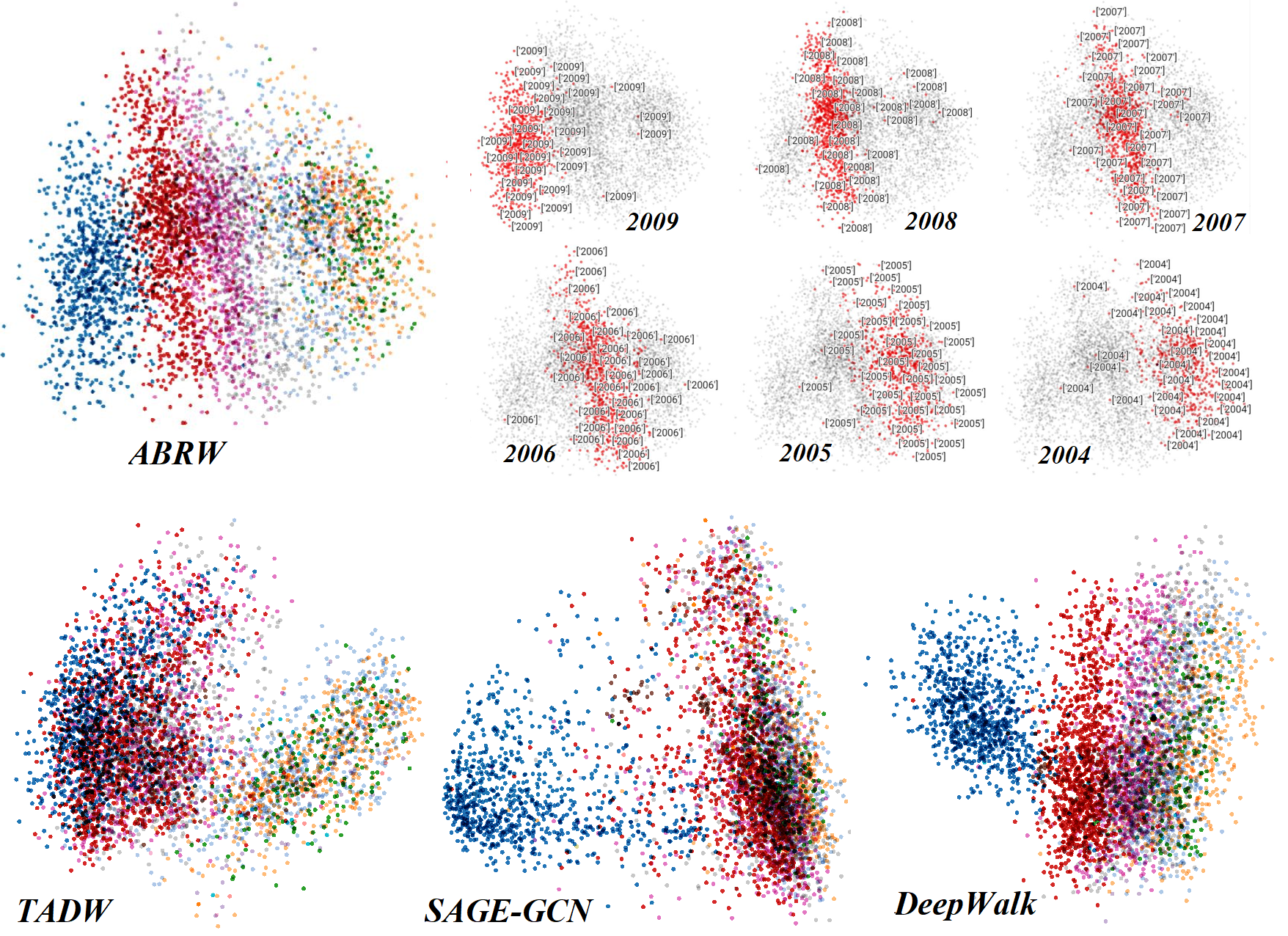}
    \caption{2D visualization of node embeddings on MIT dataset and different colors indicating different classes/years: ABRW and the highlighted years of which showing temporal trend, as well as the three competitive methods TADW, SAGE-GCN, and DeepWalk}
    \label{f4}
\end{figure}

According to the results in Figure \ref{f2} and \ref{f3}, the best four methods on MIT are selected for demonstration as shown in Figure \ref{f4}. Regarding the \textit{overview} of node embeddings, ABRW and DeepWalk show better results (more distinct clusters) than TADW and SAGE-GCN, which is consistent with the results in Figure \ref{f2} and \ref{f3}. But, DeepWalk obtains degraded results if a network becomes sparse as discussed above, \href{https://github.com/houchengbin/OpenANE}{and 2D visualization on Cora is also provided via \textit{hyperlink}.}

Besides, we observe an interesting phenomenon, namely \textit{temporal trend}, by our method as shown in the six right-hand side subfigures of the ABRW overview. It is in accordance with human common sense that students within a university have very close relationships if they graduate in the same year. And more interestingly, students also have relatively closer relationships if they graduate in nearby years (usually within 4 years). However if the year window is more than 4 years, the students graduated in 2009 have a big gap w.r.t. those students graduated in 2005 and 2004.


\subsection{Parameter Sensitivity}
We conduct link prediction ($10\%$ of links and the equal number of non-existing links serve as ground truth) to analyze the sensitivity of hyper-parameters $\alpha$ and top-k.

\begin{figure}[htbp]
    \setlength{\abovecaptionskip}{0cm}
    \setlength{\belowcaptionskip}{0cm}
    \centering
    \includegraphics[width=0.50\textwidth]{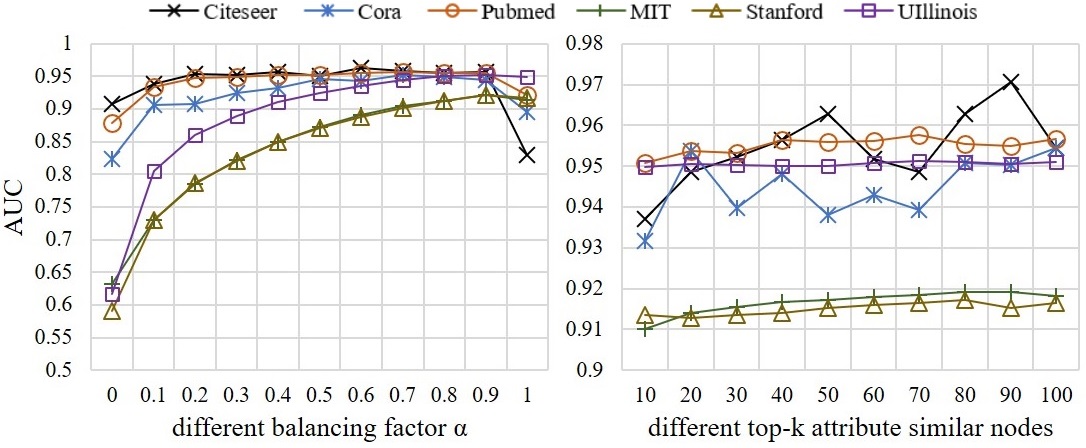}
    \caption{Sensitivity analysis of the hyper-parameters $\alpha$ and top-k}
    \label{f5}
\end{figure}

The hyper-parameter $\alpha$ is the balancing factor: $\alpha=0$ gives bias to attribute information, whereas $\alpha=1$ gives bias to structural information. For three \textit{citation networks}, they all receive comparable results, and hence, the choices of $\alpha$ can be flexible. The reason behind is that structural and attribute information are informative enough respectively, which may be observed from Figure 2 and 3 that both AttrPure and DeepWalk (with sufficient links) can obtain satisfactory results. For three \textit{social networks}, the larger $\alpha$ gives the better results. The reason behind is that structural information is more informative than attribute information, since DeepWalk (with sufficient links) significantly outperforms AttrPure.

The hyper-parameter top-k is used to preserve the top-k attribute similar nodes and remove the rest of dissimilar nodes. For three \textit{citation networks}, the results vary about $\pm 1.5\%$ w.r.t. different choices of top-k; whereas, for three \textit{social networks}, the results are quite stable. This is because the social networks have far more links than the citation networks as shown in Table 1, so that the richer structural information can better offset the inappropriate neighboring nodes added by attribute information while reconstructing the network. 


\section{Related Works}
There are two large categories of related works: 1) PNE methods by considering pure structural information, 2) ANE methods by considering  structural and attribute information. 

\textbf{Pure-structure based NE  (PNE) method}: Although this category of methods ignores attribute information, they can still obtain node embeddings based on structural information. DeepWalk \cite{perozzi2014deepwalk} applies truncated random walks to obtain node sequences which are then fed into Word2Vec model so as to embed nodes closer if they co-occur more frequently. Node2Vec \cite{grover2016node2vec} can be viewed as the extension due to it employs more flexible truncated walks to capture network structure. Besides, there are many other PNE methods such as LINE \cite{tang2015line} and HOPE \cite{ou2016asymmetric}. Nevertheless, these methods are not ideal for incomplete attributed networks, since they can only utilize incomplete structural information.

\textbf{Convex Optimization based ANE method}: TADW \cite{yang2015network} and AANE \cite{huang2017accelerated} fall into this category. They first transform structural and attribute information into two matrices respectively, and then, formulate ANE problem as a bi-convex optimization problem where the objective is to jointly minimize some distance measure between structural information matrix and embedding matrix, and also, between attribute information matrix and embedding matrix. We find that they may not converge sometimes when the structural information matrix becomes too sparse.

\textbf{Graph Convolution based ANE method}: Two representative methods of this category are GCN \cite{kipf2017semi} and graphSAGE \cite{hamilton2017inductive}. They first define node neighbors or receptive field based on network structure, and then, aggregate neighboring attribute information for further computing. These methods are not robust, as different levels of incompleteness change the definition of node neighbors, and hence their attributes to be aggregated.

\textbf{Deep Neural Networks based ANE method}: ASNE \cite{liao2018attributed} is the representative method in this category. It uses carefully-designed stacked neural networks to learn a mapping where the input and output are two node embeddings of a pair of linked nodes respectively. In other words, one link gives one training data, and obviously, incomplete structural information gives less training data, which is not desired for training a deep neural networks model.

\section{Conclusion and Future Work}
To address the challenges of embedding incomplete attributed networks, we propose an ANE method, namely Attributed Biased Random Walks (ABRW). The idea is to reconstruct a unified denser network by fusing structural and attribute information for information enhancement, and then employ a weighted random walks based network embedding method for learning node embeddings. Several experiments confirm the effectiveness of our method to incomplete attributed networks. Besides, ABRW can be viewed as a novel general framework to learn embeddings of any objects (not necessary to form a network) with multiple sources of information as long as the information can be encoded in transition matrices (what ABRW only requires), which will serve as future work.

\clearpage

\bibliographystyle{ijcai19}
\bibliography{ijcai19}

\end{document}